\documentclass[psfig]{article}
\usepackage{amssymb}
\usepackage{epsfig}

\newcommand{\be}{\begin{equation}}
\newcommand{\ee}{\end{equation}}
\newcommand{\bea}{\begin{eqnarray}}
\newcommand{\eea}{\end{eqnarray}}
\newcommand{\disp}{\displaystyle}
\title{Two Qubit Entanglement in Magnetic Chains with DM Antisymmetric Anisotropic
Exchange Interaction}

\author{Zeynep Nilhan Gurkan  and
Oktay K. Pashaev \\ \\
Department of Mathematics\\ Izmir Institute of Technology \\
Urla-Izmir, 35430, Turkey}

\begin{document}
\maketitle

\begin{abstract}

In the present paper an influence of the anisotropic antisymmetric
exchange interaction, the Dzialoshinskii-Moriya (DM) interaction, on
entanglement of two qubits in various magnetic spin models,
including the pure DM model and the most general $XYZ$ model are
studied. We find that the time evolution generated by DM interaction
can implement the SWAP gate and discuss realistic
quasi-one-dimensional magnets where it can be realized. It is shown
that inclusion of the DM interaction to any Heisenberg model
creates, when it does not exist, or strengthens, when it exists, the
entanglement. We give physical explanation of these results by
studying the ground state of the systems at T=0. Nonanalytic
dependence of the concurrence on the DM interaction and its relation
with quantum phase transition is indicated.  Our results show that
spin models with the DM coupling have some potential applications in
quantum computations and the DM interaction could be an efficient
control parameter of entanglement.
\end{abstract}
~
\newtheorem{thm}{Theorem}[subsection]
\newtheorem{cor}[thm]{Corollary}
\newtheorem{lem}[thm]{Lemma}
\newtheorem{prop}[thm]{Proposition}
\newtheorem{defn}[thm]{Definition}
\newtheorem{rem}[thm]{Remark}
\newtheorem{prf}[thm]{Proof}
\section{Introduction}
The entanglement property has been discussed at the early years of
quantum mechanics as a specifical quantum mechanical nonlocal
correlation \cite{Schrodinger}- \cite{Bell} and recently it becomes
a key point of the quantum information theory \cite{Bennet}. For
entangled subsystems the whole state vector cannot be separated into
a product of the subsystem states. This is why these subsystems are
no longer independent, even if they are far separated spatially. A
measurement on one subsystem not only gives information about the
other subsystem, but also provides possibility of manipulating it.
Therefore entanglement becomes the main tool in quantum computations
and information processing, quantum cryptography, teleportation and
etc. \cite{Angelakis}. Due to the intrinsic pairwise character of
the entanglement, entangled qubit pairs play crucial role in such
computations. It is clear that single qubit gates are unable to
generate entanglement in an $N$ qubit system, and to prepare an
entangled state one needs an inter qubit interaction, which is a two
qubit gate. The simplest two qubit interaction is described by the
Ising one between spin $1/2$ particles in the form of $J \sigma_1^z
\sigma_2^z $. More general interaction between two qubits is given
by the Heisenberg magnetic spin interaction models. These models
have been extensively studied during several decades, experimentally
in condensed matter systems \cite{wigen} and theoretically as
exactly solvable many body problems (Bethe, Baxter and others)
\cite{lieb}, \cite{baxter}. Now they become promising to realize
quantum computation and information processing, by generating
entangled qubits and constructing quantum gates \cite{zheng},
\cite{atac} in a more general context than the magnetic chains.

Recently in this way interaction of two nuclear spins having the
Heisenberg form were considered \cite{tong}. The nuclear spins from
one side are well isolated from the environment and their
decoherence time is sufficiently long. From another side nuclei with
spin $1/2$ are natural representatives of qubits in quantum
information processing, which can realize quantum computational
algorithms by using NMR \cite{yusa}, \cite{chuang}, \cite{vander}.

Very recently entanglement of two qubits \cite{Wooters2} and its
dependence on external magnetic fields, anisotropy and temperature
have been considered in several Heisenberg models: the Ising model
\cite{vedral},\cite{terzis}, \cite{childs}; the $XX$ and $XY$ models
\cite{zheng},\cite{wang3}, \cite{wang1}, \cite{kamta}, \cite{xi2},
\cite{hamieh}, \cite{sun}; the $XXX$ model \cite{arnesen}; the $XXZ$
model \cite{wang2}; and the $XYZ$ model \cite{xi1}, \cite{rigolin},
\cite{zhou}. Particularly dependence of entanglement on the type of
spin ordering, was shown, so that in the isotropic Heisenberg spin
chain (the $XXX$ model) spin states are unentangled in the
ferromagnetic case $J<0$, while for the antiferromagnetic case $J>0$
entanglement occurs for sufficiently small temperature $T<
T_c=\frac{2J}{k\ln 3}$. Significant point in the study of such
models is how to increase entanglement in situation when it already
exists or to create entanglement in situation when it does not
exist. Certainly this can be expected from a generalization of
bilinear spin-spin interaction of the Heisenberg form. Around 50
years ago explaining weak ferromagnetism of antiferromagnetic
crystals ($\alpha- Fe_2 O_3, MnCO_3 $ and $CrF_3$), has been
controversial problem for a decade, Dzialoshinskii
\cite{Dzialoshinski} from phenomenological arguments, and Moriya
\cite{Moriya} from microscopic grounds, have introduced anisotropic
antisymmetric exchange interaction, the Dzialoshinskii-Moriya (DM)
interaction, expressed by \be \vec D \cdot [\vec S_1 \times \vec
S_2].\label{DM} \ee This interaction arising from extension of the
Anderson superexchange interaction theory by including  the spin
orbit coupling effect \cite{Moriya}, is important not only for the
weak ferromagnetism but also for the spin arrangement in
antiferromagnets of low symmetry. In contrast to the Heisenberg
interaction which tends to render neighbor spins parallel, the DM
interaction has the effect of turning them perpendicular to one
another. As we will see in the present paper it turns out that such
spin arrangements are likely to increase entanglement.

In most materials with weak ferromagnetism and the DM coupling,
parameter $D$ is small compared to $J$. The values reported in the
literature range from $\frac{D}{J}\approx 0.02$ to $0.07$ (see
\cite{Aristov} and references therein). However in some compounds
the DM interaction can attain a sizeable value in comparison with
the usual symmetric superexchange $J$. Depending on compound its
value varies between $\frac{D}{J}\approx 0.05$ to $0.2$. Moreover,
recently the DM interaction was found to be present in a number of
quasi-one-dimensional magnets \cite{Pires}. Even it was found that
the compound $RbCoCl_3. 2 H_2O$ is described as a pure DM chain
\cite{Elearney}. The low-temperature magnetic behaviour of this
compound gives strong evidence that the material consists of weakly
interacting linear chains with predominant DM interaction. In
addition, study of the DM interaction influence on dynamics of the
one dimensional quantum antiferromagnet shows the big difference in
the behaviour, depending on whether the coupling $D$ is smaller or
larger than the exchange interaction $J$ \cite{Pires}. All these
results imply that a study of spin models with DM interaction could
have realistic applications. Then for applications in quantum
computations it poses the problem to find the entanglement
dependence on this interaction.

 In the present paper we study the influence of the
Dzialoshinskii-Moriya interaction on entanglement of two qubits in
all particular magnetic spin models, including the most general
$XYZ$ model. We find that in all cases, inclusion of the DM
interaction creates, when it does not exist, or strengthens, when it
exists, entanglement. For example, we show that in the case of
isotropic Heisenberg $XXX$ model discussed above, inclusion of this
term increases entanglement for antiferromagnetic case and for
sufficiently strong coupling $D> (kT sinh^{-1}e^{|J|/kT}-J^2)^{1/2}$
it creates entanglement even in ferromagnetic case. We give detailed
physical explanations of these results by studying ground state of
the system at T=0. In this state we find nonanalytic dependence of
concurrence on the DM interaction and establish its relation with
the quantum phase transition. In addition, we show that time
evolution generated by DM interaction can be implemented as the SWAP
gate. These results indicate that spin models with DM coupling have
some potential applications in quantum computations, and DM
interaction could be an efficient control parameter of entanglement.

The paper is organized as follows. In Section 2 we formulate the
general $XYZ$ model with DM coupling and find the density matrix and
eigenvalues for the concurrence. Then we consider the time evolution
and its relation with the SWAP gate. Since the concurrence
calculation depends on several parameters, in the following sections
we consider all possible particular cases from the unified point of
view. We think that such presentation is pedagogical and could be
affordable by experimentalists. In Section 3, the main properties
and entanglement of pure DM model and the relation of this model
with SWAP gate are considered. The Ising model with DM interaction
is studied in Section 4. In particular, realization of the model for
description of two nuclear spins with DM coupling and implications
for the quantum phase transitions in the presence of magnetic field
are given. In Section 5 we consider the $XY$ model and its
particular reductions to the $XX$ case, and to the Ising model. We
show that inclusion of the transverse magnetic field leads to the
different behaviour of concurrence $C_{12}$ for the undercritical
and the overcritical couplings. For $T=0$ the nonanalytic behaviour
for $C_{12}(D)$ is found. The $XXX$ Heisenberg model is subject of
Section 6. Section 7 is devoted to the $XXZ$ model, where the
influence of DM coupling and magnetic field on the concurrence and
the quantum phase transitions are studied. In Section 8 we study
$XYZ$ model in both antiferromagnetic and ferromagnetic cases, with
inclusion of the DM coupling. The nonanalytic behaviour at $T=0$ is
found. In Conclusions several implications for future studies are
discussed.

\section{$XYZ$ Heisenberg Model}

We start our consideration with the most general $XYZ$ model, by
inclusion of homogeneous $B$ and nonhomogeneous $b$ magnetic fields,
and choosing the DM interaction (\ref{DM}) in the form $
\frac{\vec{D}}{2}= \frac{D}{2} \cdot \vec z $. Then for two qubits
we have Hamiltonian
 \be H = \frac{1}{2} [J_x \, \sigma^x_1
\sigma^x_2 + J_y \,\, \sigma^y_1 \sigma^y_2 + J_z \,\, \sigma^z_1
\sigma^z_2 + B_+ \, \sigma^z_1 +B_-\, \sigma^z_2 + D (\sigma^x_1
\sigma^y_2 - \sigma^y_1 \sigma^x_2 ) ]\label{Hamiltonian1} \ee
  where $B_+ \equiv B+b, B_- \equiv B-b $ and  $\sigma_i^x, \sigma_i^y, \sigma_i^z $, $i=1,2$
  denote Pauli matrices related with the first and the second qubits.
\subsection{Eigenvalues and Eigenvectors}
To study the thermal entanglement in this system firstly we need to
obtain all eigenvalues and eigenstates of the Hamiltonian
(\ref{Hamiltonian1}): $ H|\Psi_i \rangle = E_i | \Psi_i \rangle ,
\,\,\,\, i= 1,2,3,4. \label{Hamiltonianeqn}$ Simple calculations
show that the energy levels are: \be E_{1,2}= \frac{J_z}{2}\mp
\mu,\,\,\, E_{3,4}= -\frac{J_z}{2}\mp \nu  \label{eigenvalues1} \ee
where  $\mu \equiv \sqrt{B^2 + J_-^2}$, $\nu \equiv \sqrt{b^2+
J_+^2+D^2}$, $J_\pm \equiv \frac{J_x \pm J_y}{2}$, and the
corresponding wave functions are {\small
\begin{equation} |\Psi_{1,2}\rangle = \frac{1}{\sqrt{2\mu
(\mu\pm B)}} \left[
     \begin{array}{c}
           J_- \\
           0 \\
           0 \\
           -(B\pm\mu) \\
         \end{array}
       \right] ,\,\,\,\
  |\Psi_{3,4}\rangle =\frac{1}{\sqrt{2\nu(\nu \mp b)}} \left[
    \begin{array}{c}
           0\\
          (b\mp\nu)\\
          J_+-iD \\
          0 \\
         \end{array}
        \right]\label{psi3-4} \end{equation} }

For $B=0,\, b=0 ,\, D=0$ these wave functions reduce to the
maximally entangled Bell states {\small \bea |\Psi_{2,1}\rangle
\longrightarrow |B_{0,3}\rangle &=& \frac{1}{\sqrt{2}}\,(|00 \rangle
\pm |11 \rangle)
          \\
|\Psi_{4,3}\rangle \longrightarrow |B_{1,2}\rangle &=&
\frac{1}{\sqrt{2}}\,(|01 \rangle \pm |10 \rangle \eea }

\subsection{Time Evolution of States and SWAP Gate}

Here we like to show the direct relationship between our spin model
and quantum gates. For this we consider the evolution operator \be
U(t)= \exp [ -\frac{i}{\hbar} H t]\ee determined by two qubit
Hamiltonian (\ref{Hamiltonian1}) of $XYZ$ model with DM coupling,
$B=0$, $b=0$. Then evolution of the standard basis is given by \bea
| 00 \rangle & \rightarrow & e^{\frac{-i J_z t}{2 \hbar}} \left[
\cos \frac{t J_{-}}{\hbar} |00\rangle -i \sin \frac{t J_{-}}{\hbar}
|11\rangle \right] \label{00},\\
| 11 \rangle & \rightarrow & e^{\frac{-i J_z t}{2 \hbar}} \left[\cos
\frac{t J_{-}}{\hbar} |11\rangle - i \sin \frac{t J_{-}}{\hbar}
|00\rangle \right] \label{01},\\
| 01 \rangle & \rightarrow & e^{\frac{i J_z t}{2 \hbar}} \left[\cos
\frac{t \nu }{\hbar} |01\rangle - i \frac{J_+ - i D}{\nu} \sin
\frac{t \nu }{\hbar}
|10\rangle \right]\label{10}, \\
| 10 \rangle & \rightarrow & e^{\frac{i J_z t}{2 \hbar}} \left[\cos
\frac{t \nu }{\hbar} |10\rangle - i \frac{J_+ + i D}{\nu} \sin
\frac{t \nu }{\hbar} |01\rangle \right] \label{11}
 \eea
where $\nu= \sqrt{J_+^2+ D^2}$. In particular cases, discussed in
the next section, this evolution can implement the SWAP gate at time
$t=\hbar \pi/2\nu$.

\subsection{Density Matrix and Concurrence}
State of the system at thermal equilibrium is determined by the
 density matrix
 \be \rho(T)= \frac{e^{-H/ k T}}{Tr[e^{-H /
k T}]}=\frac{e^{-H/ k T}}{Z} , \label{densitymatrix}\ee where
$Z=Tr[e^{-H/kT}]$ is the partition function, $k$ is the Boltzmann
constant and $T$ is the temperature. Then by exponentiation of
Hamiltonian (\ref{Hamiltonian1})
 we find
 {\small \be e^{-H / k T} = \left[
         \begin{array}{cccc}
            A_{11} & 0 & 0 & A_{14} \\
           0 &   A_{22} & A_{23} & 0 \\
           0 &  A_{32} &  A_{33} & 0 \\
           A_{41} & 0 & 0 &  A_{44}  \\
         \end{array}
       \right]
\label{rho} \ee}
   where
\small{ \bea A_{11}&=& e^{\frac{- J_z}{2kT}}\left[ \cosh
\frac{\mu}{kT}-\frac{B}{\mu}\sinh \frac{\mu}{kT}\right] \nonumber\\
A_{44} &=& e^{-\frac{J_z}{2kT}}\left[ \cosh
\frac{\mu}{kT}+\frac{B}{\mu}\sinh \frac{\mu}{kT}\right] \\
A_{14}&=& - e^{-\frac{J_z}{2kT}}\frac{J_-}{\mu} \sinh
\frac{\mu}{kT}\nonumber\\
 A_{41}&=&-e^{-\frac{J_z}{2kT}}\frac{J_-}{\mu} \sinh
\frac{\mu}{kT}\nonumber\\
A_{22}&=& e^{\frac{J_z}{2kT}}\left[ \cosh
\frac{\nu}{kT}-\frac{b}{\nu}\sinh \frac{\nu}{kT}\right]\nonumber\\
A_{33}&=& e^{\frac{J_z}{2kT}}\left[ \cosh
\frac{\nu}{kT}+\frac{b}{\nu}\sinh \frac{\nu}{kT}\right] \nonumber\\
A_{23}&=& -e^{\frac{J_z}{2kT}}\frac{J_++iD}{\nu}\sinh \frac{\nu}{kT}
\nonumber\\
A_{32}&=& -e^{\frac{J_z}{2kT}}\frac{J_+-iD}{\nu}\sinh \frac{\nu}{kT}
 \eea and \be Z=Tr[e^{-H/kT}]= 2 \left[ e^{\frac{-J_z}{2kT}} \cosh
\frac{\mu}{kT}+ e^{\frac{J_z}{2kT}} \cosh \frac{\nu}{kT}
\right].\nonumber\ee}As $\rho(T)$ represents a thermal state, the
entanglement in this state is called the \textbf{\emph{thermal
entanglement}}. The degree of entanglement could be characterized by
the concurrence $C_{12}$, which is defined as \cite{Wooters2},
\cite{Wooters1} \be C_{12}= max
\{\lambda_1-\lambda_2-\lambda_3-\lambda_4, 0\} \label{concurrence}
,\ee where $\lambda_1\geq \lambda_2 \geq \lambda_3 \geq \lambda_4
> 0$ are the ordered square roots of eigenvalues of the operator
\be \rho_{12}= \rho (\sigma^y \otimes \sigma^y) \rho^{*}(\sigma^y
\otimes \sigma^y) .\ee  The concurrence is bounded function $0 \le
C_{12} \le 1$, so that when $C_{12}=0$, the states are unentangled,
while for $C_{12}=1$, the states are maximally entangled.

For the general Hamiltonian (\ref{Hamiltonian1}) we find :

{\small \begin{eqnarray} \lambda_{1,2} &=&
\frac{e^{\frac{-J_z}{2kT}}}{Z} \left|\sqrt{1+
\frac{J_-^2}{\mu^2}\sinh^2 \frac{\mu}{kT}}\mp
\frac{J_-}{\mu} \sinh \frac{\mu}{kT}\right|\\
\lambda_{3,4}&=& \frac{ e^{\frac{J_z}{2kT}}}{Z} \left|\sqrt{1+
\frac{J_+^2+D^2}{\nu^2}\sinh^2 \frac{\nu}{kT}}\mp
\frac{\sqrt{J_+^2+D^2}}{\nu} \sinh \frac{\nu}{kT}\right|. \nonumber
\end{eqnarray}}
Then, to calculate the concurrence we need to order these
eigenvalues. Since they depend on several parameters, before
studying the most general case, it is useful to treat all particular
cases separately to clarify the influence of the DM coupling on the
entanglement. Starting from pure DM model we study
 various Heisenberg models, including the general $XYZ$ case.

Before this, we like just to stress here the general observation on
the concurrence (\ref{concurrence}). If the biggest eigenvalue say
$\lambda_1$ is degenerate, then its positive contribution would be
compensated by the another degenerate one, so that $C_{12}=0$ and
states are always unentangled. We will encounter this situation in
several cases and it has a simple physical explanation. The
degenerate biggest eigenvalues of the density matrix correspond to
the minimal values of the energy, so that the ground state of the
system becomes degenerate and no entanglement occurs.

\section{Pure DM Model}
\subsubsection{Main Characteristics of DM Model}
As we discussed in introduction some realistic quasi-one dimensional
compounds with predominance of DM interaction can be described as a
pure DM chain \cite{Elearney}. Here we consider the main
characteristic properties of the DM coupling between two qubits and
its influence on the entanglement. If in Hamiltonian
(\ref{Hamiltonian1}) we put $J_x=J_y=J_z=0$ and $B=b=0$ then the
model is determined completely by the DM term (\ref{DM}). In this
case the first two eigenstates become degenerate $E_1=E_2=0$ and
$E_{3,4}= \pm D$. For definiteness we choose $D>0$, then for $T=0$
the ground state of the system with energy $E_4=-D$ is an entangled
state $|10\rangle - i |01\rangle $. When temperature increases this
state becomes mixed with the higher states and entanglement
decreases. But for sufficiently large value of $D$ the ground state
can be alienated so that entanglement increases. This shows that for
a given $D$ there exists $kT_c= D / \ln (1+ \sqrt{2})$ so that for
the under critical case $T<T_c$ the states become entangled and the
concurrence is $C_{12}= \frac{\sinh \frac{D}{kT}-1}{\cosh
\frac{D}{kT}+1}$ (See Fig.1). For $T=0$ the concurrence $C_{12}=1$
and the ground state is maximally entangled.

\begin{figure}[htbp]
\centerline{\epsfig{figure=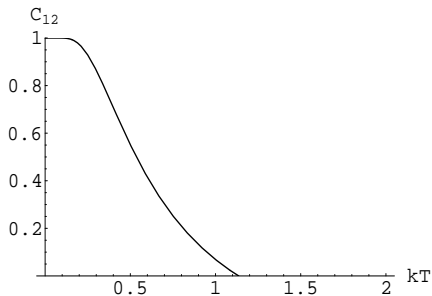
,height=4cm,width=6cm}} \caption{Concurrence versus temperature for
$D=1$ and $T_c= 1.136$}
\end{figure}

\subsubsection{DM Model and SWAP Gate}
The time evolution in pure DM model from one side is related with
the SWAP gate, from another side can create maximally entangled
states.
 In this case according to
(\ref{00})-(\ref{11}) for time evolution we have \bea
U(\frac{\pi \hbar}{2 D})| 00 \rangle & =& |00\rangle ,\,\,\,\,\,\,\,\,U(\frac{\pi \hbar}{2 D})| 11 \rangle = |11\rangle \label{U00}\\
U(\frac{\pi \hbar}{2 D})| 01 \rangle &=& -|10\rangle
,\,\,\,\,\,\,\,\, U(\frac{\pi \hbar}{2 D})| 10 \rangle = -|01\rangle
\label{U10}. \eea Therefore we can see that the operator
$U(\frac{\pi \hbar}{2 D})$ acts as the SWAP gate. Moreover at time
$t= \pi \hbar / 4 D$ the states $| 01 \rangle$ and $| 10 \rangle$
becomes maximally entangled Bell states. \bea U(\frac{\pi \hbar}{4
D})| 01 \rangle & =& \frac{1}{\sqrt{2}} (|01\rangle- |10\rangle)
\\
U(\frac{\pi \hbar}{4 D})| 10 \rangle & =& \frac{1}{\sqrt{2}}
(|10\rangle+ |01\rangle) \eea

\section{Ising Model}
For $J_x=J_y=0$, $J_z \neq 0$ and $B=b=0, D=0$ the Hamiltonian
(\ref{Hamiltonian1}) describes the Ising model . It was observed
before that for pure Ising model in both the antiferromagnetic
,$J_z>0$, and the ferromagnetic cases, $J_z<0$, the concurrence is
zero and the states are always unentangled \cite{vedral},
\cite{terzis}, \cite{childs}. The physical insight of such behavior
is easy to understand. When $J_-=J_+=0$ the density matrix $\rho$
(\ref{densitymatrix}) is diagonal in the standard basis which
implies the absence of quantum correlations. Despite of having four
maximally entangled states as the eigenvectors, the states
$|\Psi_{1,2}\rangle$ and $|\Psi_{3,4}\rangle$ are degenerated, so
that the Ising thermal state has no entanglement. The situation does
not change if one includes homogeneous $B$ or nonhomogeneous $b$
magnetic fields, because the density matrix $\rho$ is still diagonal
and no entanglement occurs .

\subsection{Ising Model with DM Coupling ($B=0, b=0, D\neq 0 $)}
In contrast to magnetic fields, which does not create entanglement,
inclusion of the DM coupling contributes to the nondiagonal elements
of $\rho$ and creates entanglement.
\subsubsection{Antiferromagnetic Case ($J_z>0$)}
In this case addition of the DM coupling to the Ising model splits
the degenerate ground state with $E_3= E_4= -\frac{J_z}{2}$ so that
it becomes a singlet with $E_3= -\frac{|J_z|}{2}- D$, for $D>0$ or
$E_4= -\frac{|J_z|}{2}+ D$, for $D<0$. At $T=0$ this leads to the
maximally entangled state with $C_{12}=1$. When temperature
increases the maximally entangled ground state becomes mixed with
the higher eigenstates and the entanglement decreases. However, for
a given temperature by increasing the coupling $D>D_c$, where
$D_{c}= kT \sinh^{-1} e^{-J_z/ kT}$, we can decrease this mixture
and increase entanglement, so that the concurrence is \be C_{12} =
\frac{\sinh \frac{|D|}{kT}-e^{-J_z/kT}}{\cosh\frac{|D|}{kT}+
e^{-J_z/ kT}}.
 \label{concAFIsingDM} \ee

 \subsubsection{Ferromagnetic Case ($J_z<0$)}
In this case the ground state for small $D$ at $T=0$ is also a
doublet and no entanglement occurs. However, with growing $D$ the
eigenstate $E_3= \frac{|J_z|}{2}-D$ is lowering so that at critical
value $D_c=|J_z|$ the ground state becomes triplet. When $D>D_c$ the
ground state $E_3$ is maximally entangled singlet. With growing
temperature, a mixture of this state with the higher states
decreases entanglement. For given temperature $T$, there exist the
critical value $D_c=|J_z|+ \frac{kT}{2}\ln (1+ e^{{-2|J_z|}/{kT}})$
so that for $D>D_c$ the concurrence is {\small \be C_{12} =
\frac{\sinh \frac{|D|}{kT}-e^{|J_z|/kT}}{\cosh\frac{|D|}{kT}+
e^{|J_z|/ kT}}. \label{96}\ee }

Comparison of (\ref{concAFIsingDM}) and (\ref{96}) shows that in the
antiferromagnetic case the states can be entangled more easily than
in the ferromagnetic one.

 \subsection{Ising Model for Two Nuclear Spins with DM Coupling}
As an application of the above calculations here we discuss
entanglement of two nuclear spins. Recently two nuclear spins were
considered in a model with weak Heisenberg type interaction in a
constant longitudinal magnetic field along $z$ direction
\cite{tong}\bea H&=& H_z+
H_{xy}\\
H_z&=&-\frac{1}{2}(\omega_1 \sigma_1^z+ \omega_2 \sigma_2^z+
J \sigma_1^z \sigma_2^z) \label{Hz}\\
H_{xy}&=& -\frac{1}{2} (J \sigma_1^x \sigma_2^x+ J \sigma_1^y
\sigma_2^y) \eea where the isotropic form for the spin coupling $J$
is assumed, and $\omega_{1,2}\equiv(B\mp b)$ are the Larmor
frequencies of two nuclear spins, $\hbar=1$. In the experiments, two
different nuclear spins are selected, $\omega_1 \neq \omega_2$ (we
assume $\omega_1 >\omega_2$), and the longitudinal constant magnetic
field is in the order of $1THz$, so that $\omega_1, \omega_2$ are
much larger than $J$ and $\eta = \frac{J}{(\omega_1 -\omega_2)}\ll
1$. $H_{xy}$ is non-diagonal in $\sigma_z$ representation and due to
quantum fluctuations of order $\eta^2$, can be ignored. Thus, the
Ising part $H_z$ of the Hamiltonian is a well precise approximation
\cite{tong}. However as we have seen above, for the Ising model with
external magnetic fields no entanglement occurs, this is why two
nuclear spins in this model are unentangled for any $\omega_1$ and
$\omega_2$. From another side, as follows from our consideration in
Sec.4.1 the addition of an interaction between qubits in the form of
the DM coupling could make them entangled. Now by adding the DM
interaction to two nuclear spin Hamiltonian (\ref{Hz}) we get the
Ising model with homogeneous magnetic field $B$, nonhomogeneous
magnetic field $b$ and the DM interaction $D$. In the
antiferromagnetic and the ferromagnetic cases, when $J_z= \pm |J_z|$
respectively, for sufficiently strong $D> D_c$, where
$\frac{D_c}{\sqrt{D_c^2+ b^2}} \sinh{\frac{\sqrt{D_c^2+b^2}}{kT}}=
e^{\mp\frac{|J_z|}{kT}}$, the states become entangled and the
concurrence is \be C_{12}= \frac{\frac{D}{\nu}\sinh \frac{\nu}{kT}
-e^{\mp\frac{|J_z|}{kT}}}{\cosh\frac{\nu}{kT} + \cosh\frac{B}{kT}
e^{\mp\frac{|J_z|}{kT}}} \ee where $B=(\omega_1+\omega_2)/2$,
$b=(\omega_1-\omega_2)/2$ and $\nu=
\sqrt{\frac{(\omega_2-\omega_1)^2}{4}+D^2}$. It is worth to note
that the homogeneous magnetic field $B$ does not change critical
value for the entanglement, but could change level of the
entanglement. Moreover, increasing magnetic field, decreases value
of the entanglement. It turns out that for the system at $T=0$, the
concurrence becomes nonanalytic when $D=D_c$

\be C_{12}=\left\{%
\begin{array}{ll}
  \frac{D}{\nu} , & \hbox{$ \nu> B\mp|J_z|$;} \\ \\
 \frac{D}{2\nu}, & \hbox{$\nu= B\mp |J_z|$;}\\ \\
  0   , & \hbox{$\nu< B\mp |J_z|$,}
\end{array}%
\right.     \ee that implies quantum phase transitions at the
critical value $D_c= (B\mp |J_z|)^2-b^2$.

\subsubsection{Ising Model with DM Coupling and SWAP Gate}

If $J_x=J_y=0$ but $J_z$ and $D$ are nonvanishing and related  by
$J_z= 8n\, D,\,\,\, (n= \pm 1, \pm 2 ...)$, then again like in
Sec.3.02 the evolution operator $U(\pi \hbar / 2 D)$ acts as the
SWAP gate. Our consideration shows that the Ising model, which was
derived in several physical situations for interaction of qubits,
with addition of the DM coupling, from one side leads to
entanglement of states, from another side it can model the SWAP gate
(\ref{U00}), (\ref{U10}). This result shows that the Ising model
with DM coupling have some potential applications in quantum
computations.


\section{$XY$ Heisenberg Model}
In the pure $XY$ Heisenberg Model $J_z=0, J_x \neq J_y$ and $B=0,
b=0, D=0 $ in (\ref{Hamiltonian1}), for the antiferromagnetic case
$J_x>0$, $J_y>0$ the ordered eigenvalues are
        $\lambda_3>\lambda_1> \lambda_2 > \lambda_4$
and for $\sinh \frac{J_+}{kT} > \cosh \frac{J_-}{kT}$ the
entanglement occurs with  $C_{12}= \frac{\sinh \frac{J_+}{kT} -
\cosh \frac{J_-}{kT}}{\cosh \frac{J_-}{kT}+ \cosh \frac{J_+}{kT}} $.
In the ferromagnetic case $J_x<0$, $J_y<0$ the entanglement occurs
when
 $\sinh
\frac{|J_-|}{kT} > \cosh \frac{J_+}{kT}$  with the concurrence
\cite{wang1}, \cite{hamieh}, \cite{kamta}, \cite{sun} \be C_{12}=
\frac{\sinh \frac{|J_+|}{kT} - \cosh \frac{J_-}{kT}}{\cosh
\frac{|J_-|}{kT}+ \cosh \frac{J_+}{kT}}. \ee For the particular case
of pure $XX$ model, when  $J_x=J_y\equiv J$, in both
antiferromagnetic and ferromagnetic cases the states become
entangled at sufficiently small temperature \be T< T_c =
\frac{|J|}{k \,sinh^{-1}1}\label{criticalXX}.\ee   As was shown in
\cite{zheng}, \cite{atac},\cite{wang3}, \cite{xi1} inclusion of the
magnetic field does not change this critical temperature.

\subsection{$XY$ Heisenberg Model with DM Coupling ($B = 0, b = 0 , D
\neq 0$)}

By addition of the DM coupling eigenvalues become $ \lambda_{1,2}=
\frac{e^{\pm J_-/kT}}{Z},\,\,\, \lambda_{3,4}= \frac{e^{\pm
\sqrt{J_+^2+D^2}/kT}}{Z}$ where $ Z= 2\left[\cosh \frac{|J_-|}{kT}+
\cosh \frac{\sqrt{J_+^2+ D^2}}{kT}\right].$

In the antiferromagnetic case for any temperature $T$ we can adjust
sufficiently strong DM coupling $D$ so that for $\sinh
\frac{\sqrt{J_+^2+D^2}}{kT} > \cosh
    \frac{J_-}{kT}$ the entanglement occurs with concurrence \be C_{12}=  \frac{\sinh \frac{\sqrt{J_+^2+D^2}}{kT} - \cosh
\frac{J_-}{kT}}{\cosh \frac{\sqrt{J_+^2+D^2}}{kT}+ \cosh
\frac{J_-}{kT}}.\ee The ferromagnetic case gives the same result as
the antiferromagnetic one. The comparison with pure XY model
shows that the level of entanglement increases with growing coupling
D.

In particular case $J_x=J_y\equiv J$, the ordered eigenvalues are $
\lambda_4=\frac{e^{\nu/kT}}{Z}> \lambda_3= \frac{e^{-\nu/kT}}{Z}
>\lambda_{1,2}= \frac{1}{Z}$, where $\nu= \sqrt{J^2+ D^2}$ and $Z=
2(1+ \cosh \frac{\nu}{kT})$. Then the entanglement occurs when
$\sinh\frac{\nu}{kT}>1$ and the concurrence is $\disp
C_{12}=\frac{\sinh\frac{\nu}{kT}-1}{\cosh \frac{\nu}{kT}+1} $.
Comparison with the pure XX model (\ref{criticalXX}) shows that the
critical temperature  \be T_c = \frac{\sqrt{J^2 + D^2}}{k\,
sinh^{-1}1}\ee in this case increases with growing $D$. For $D=0$
$|\Psi_3\rangle$ in (\ref{psi3-4}) is the ground state with
eigenvalue $E_3=-|J_+|$, which is maximally entangled Bell state, so
that the concurrence $C_{12}=1$. As $T$ increases the concurrence
decreases due to the mixing of other states with this maximally
entangled one\footnote{In ref. \cite{wang2} entanglement in $XX$
model with DM coupling was derived but not in the general $XXZ$ case
as it is claimed in the paper.} .

\subsection{Ising Model in Transverse Magnetic Field}
As a particular case of the general $XY$ model now we consider the
transverse Ising model, when $J_y=0$, with external magnetic field
$B$ in $z-$ direction \cite{kamta}, and with addition of DM
interaction: \be H= \frac{1}{2}[J_x (\sigma^x_1 \sigma^x_2) + B
(\sigma^z_1 + \sigma^z_2)+ D(\sigma^x_1 \sigma^y_2 - \sigma^y_1
\sigma^x_2)]. \ee The corresponding eigenvalues and the partition
function $Z$ can be written as follows  \bea \lambda_{1,2}&=&
\frac{1}{Z} \left| \sqrt{ 1
+ \frac{J^2}{B^2+J^2}\sinh^2 \frac{\sqrt{B^2+J^2}}{kT}} \mp \frac{J}{\sqrt{B^2+J^2}} \sinh{\frac{\sqrt{B^2+J^2}}{kT}}\right|\\
\lambda_{3,4}&=& \frac{1}{Z} e^{\mp \frac{\sqrt{J^2+D^2}}{kT}},
 \eea
\be Z= 2 \left[\cosh \frac{\sqrt{B^2+J^2}}{kT}+ \cosh
\frac{\sqrt{D^2+J^2}}{kT}\right].\ee To find the maximal eigenvalue
we
 compare the difference of $\lambda_4$ and $\lambda_2$ as a
function of $B, D$ and $T$, $ \lambda_4-\lambda_2 \equiv f(B, D,
T)$: \be f= e^{\frac{\sqrt{J^2+D^2}}{kT}}- \sqrt{ 1 +
\frac{J^2}{B^2+J^2}\sinh^2 \frac{\sqrt{B^2+J^2}}{kT}}-
\frac{J}{\sqrt{B^2+J^2}}
\sinh{\frac{\sqrt{B^2+J^2}}{kT}}\label{compare} \ee When $f(B, D,
T)=0$ we find the critical $D=D_c(B,T)$ as \bea D_c(B,T)= \hskip10cm\\
\sqrt{-J^2+ T^2 \left(\ln \left[ \sqrt{1+ \frac{J^2}{B^2+J^2}
\sinh^2 \frac{\sqrt{B^2+J^2}}{kT}}+
\frac{J}{\sqrt{B^2+J^2}}\sinh{\frac{\sqrt{B^2+J^2}}{kT}}
\right]\right)^2}. \nonumber\eea In Fig.2 we plot $D_c$ as a
function of $T$ for different values of magnetic field $B=0.05, 0.5,
0.7, 1$ ($J=1, k=1$). The 3D plot of $D_c$ as a function of $B$ and
$T$ for the same values of parameters is given in Fig.3.
\begin{figure}[htbp]
\centerline{\epsfig{figure=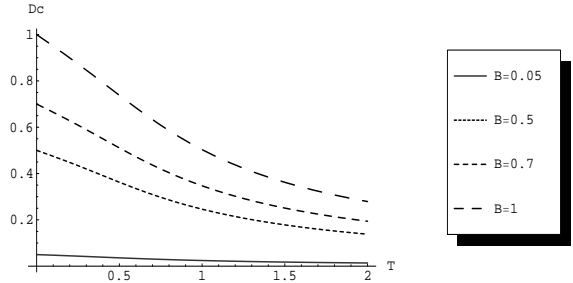
,height=4cm,width=8cm}} \caption{$D_c$ versus $T$ for $B= 0.05, 0.5,
0.7, 1$}
\end{figure}

\begin{figure}[htbp]
\centerline{\epsfig{figure=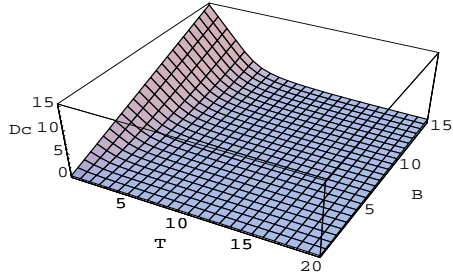
,height=4cm,width=6cm}} \caption{3D plot $D_c$ versus $B$ and $T$}
\end{figure}

For critical $D=D_c$, the eigenvalues are degenerate
$\lambda_2=\lambda_4$ and as a result the concurrence
$C_{12}(B,D_c,T)=0 .$ However the value of concurrence is different
for the under critical and the over critical cases. In under
critical case when $D<D_c$ the maximal eigenvalue is $\lambda_2$ and
for the concurrence we have

\be C_{12}= max \{ \frac{\frac{J}{\sqrt{B^2+J^2}}
\sinh{\frac{\sqrt{B^2+J^2}}{kT}} - \cosh \frac{\sqrt{D^2+J^2}}{kT}
}{\cosh \frac{\sqrt{B^2+J^2}}{kT}+\cosh \frac{\sqrt{D^2+J^2}}{kT}} ,
0 \},\ee while in the over critical case, when $D>D_c$,  the maximum
eigenvalue is $\lambda_4$ and the concurrence is \be C_{12}= max \{
\frac{\sinh{\frac{\sqrt{D^2+J^2}}{kT}- \sqrt{ 1 +
\frac{J^2}{B^2+J^2}\sinh^2 \frac{\sqrt{B^2+J^2}}{kT}} } }{\cosh
\frac{\sqrt{B^2+J^2}}{kT}+\cosh \frac{\sqrt{D^2+J^2}}{kT}},0 \}.\ee
In pure Ising model when $B=0$ and $D=0$ as we can see from
(\ref{compare}) we have $f(0,0,T)=0$ and no entanglement occurs. But
as reported in \cite{kamta} an addition of the transverse magnetic
field to the Ising model could create entanglement. Now we can
generalize these results by analyzing in addition the influence of
DM interaction on entanglement in the Ising model with the magnetic
field. When $B=0$ the addition of solely DM term creates
entanglement at sufficiently strong $D$, and this value of $D$
becomes bigger for higher temperatures. If we have both terms $B
\neq 0$ and $D \neq 0$, then with increasing $D$ the behavior of
entanglement becomes nontrivial. In Figs. 4.a, 4.b, 4.c we show
behavior of entanglement as a function of $D$ for different
temperatures. When $T=0$ entanglement is nonanalytic function of
$D$, given by the step function

\be C_{12}(D)=\left\{%
\begin{array}{ll}
    \frac{J}{\sqrt{J^2+B^2}}, & \hbox{$D<D_c$ ;} \\ \\
    0, & \hbox{$D=D_c$ ;} \\ \\
    1, & \hbox{$D>D_c$ .} \\
\end{array}%
\right. \ee  where $D_c=B$ (see Fig. 4-a). This nonanalytic behavior
signals on the quantum phase transition \cite{Sachdev} appearing at
$D=D_c=1$. In Fig. 4-b  at temperature $T=0.5$ the entanglement as a
function of $D$ decreases down to zero and at $D_c \approx 0.75$
reaches its nondifferentiable minima. After this it increases
monotonically with growing $D$. For higher temperature $T=1$ in Fig.
4-c, the entanglement is zero until $D$ becomes sufficiently strong
at $D=D_c$, where entanglement appears and monotonically grows with
growing $D$.
\begin{figure}[htbp]
\centerline{\epsfig{figure=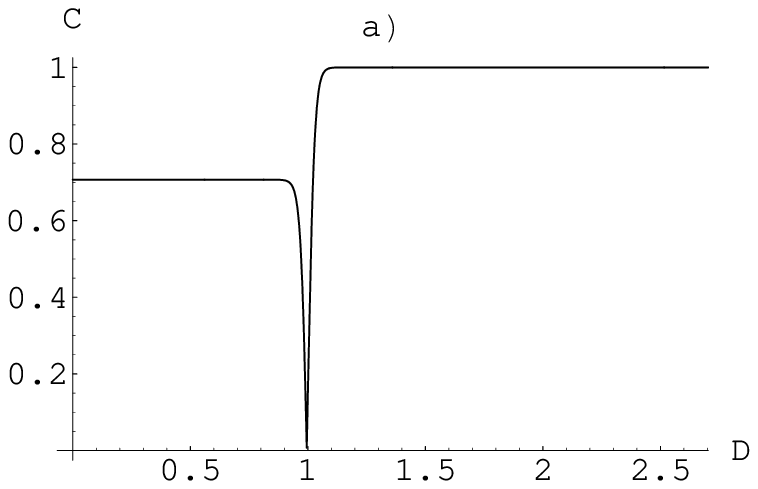
,height=2.7cm,width=2.7cm}
\epsfig{figure=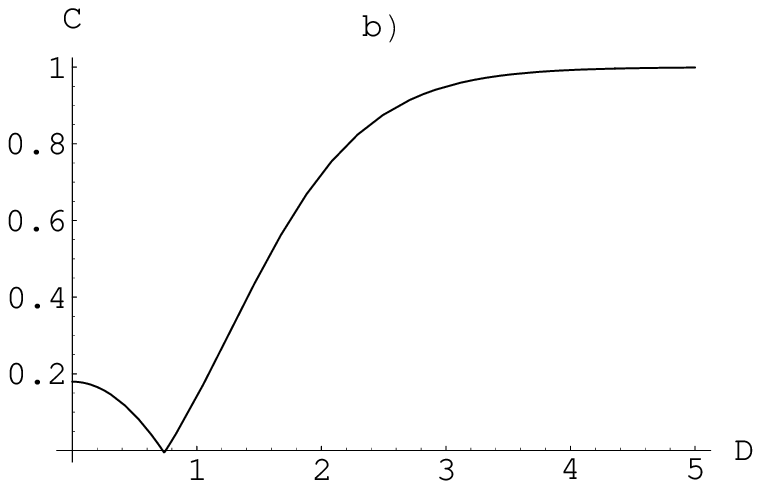,height=2.7cm,width=2.7cm}
\epsfig{figure=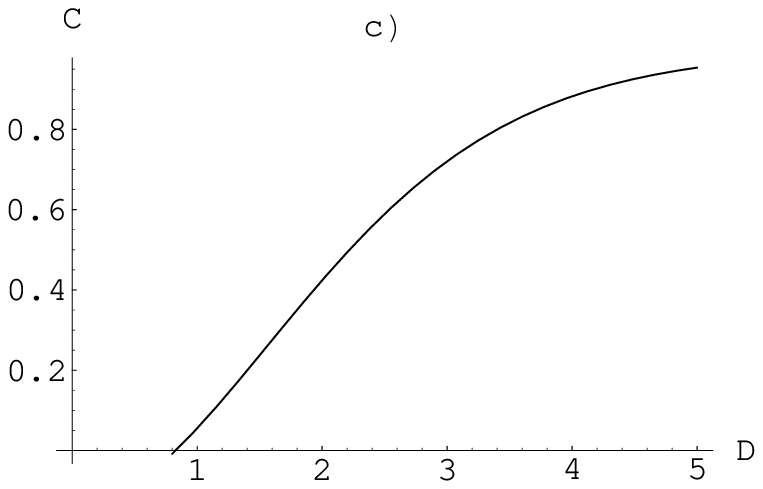,height=2.7cm,width=2.7cm} }
\caption{Concurrence of Ising model in transverse magnetic field
versus D, when $B=1$ and $T=0.01,\,0.5,\,1$}
\end{figure}

\section{$XXX$ Heisenberg Model}
In pure $XXX$ model $J_x=J_y=J_z \equiv J$ and $B=b=D=0 $ in
(\ref{Hamiltonian1}), entanglement behavior for the ferromagnetic
and the antiferromagnetic cases is different. In the spectrum of the
model we have three degenerate eigenstates with eigenvalue $J/2$ and
one eigenstate with eigenvalue $-3J/2$. It was observed before
\cite{arnesen} that for the ferromagnetic case ($J<0$) the
concurrence is zero and the states are always unentangled. It
happens because when $J<0$, the ground state of the system is an
equal mixture of the triplet states with energy, $E_1=E_2=E_4=-
\frac{|J|}{2}$. The density matrix $\rho$ is diagonal and inclusion
of magnetic field does not change the result. Increasing temperature
$T$ just increases the singlet mixture with the triplet, which can
only decrease entanglement \cite{arnesen}, \cite{nielsen}. The
situation is different for the antiferromagnetic case when $J>0$. In
this case the ground state is the maximally entangled singlet state
with $E_3= -\frac{3 J}{2}$, so that the concurrence $C_{12}=1$ at
$T=0$. It decreases with $T$ due to mixing of the triplet higher
states with the singlet ground state. For a given coupling constant
$J$ entanglement occurs at temperature $T< \frac{2J}{k\ln 3}$
\cite{wang2}.

\subsection{$XXX$ Heisenberg Model with DM Coupling ($B = 0, b
= 0, D\neq 0$)} Now by adding DM coupling for the antiferromagnetic
and the ferromagnetic cases, for $J= \pm |J|$ respectively, for a
given temperature $T$ the entanglement occurs when $ D> D_c=
\sqrt{(kT \sinh^{-1}e^{\mp|J|/kT})^2-J^2}$ with the concurrence
{\small \be C_{12}= \frac{\sinh \frac{\sqrt{J^2+D^2}}{kT}-
e^{\mp|J|/kT}}{e^{\mp|J|/kT}+ \cosh \frac{\sqrt{J^2+D^2}}{kT}}. \ee
}

 As we can see inclusion of the DM coupling, in the $XXX$ model,
increases entanglement in the antiferromagnetic case and creates
entanglement even in the ferromagnetic case. This can be explained
if we consider the eigenvalues of our Hamiltonian varying with $D$.

For the antiferromagnetic case the ground state of the system
remains singlet with energy $E_3= -\frac{|J|}{2}- \sqrt{J^2+D^2}$,
while from degenerate excited triplet state one of the energy levels
$E_4= -\frac{|J|}{2}+ \sqrt{J^2+D^2}$ is splitting up. With
increasing coupling $D$ the gap between ground state and the first
excited doublet state is increasing, this is why the system becomes
more entangled.

In the ferromagnetic case, from unentangled triplet ground state one
of the states splits with the energy $E_3= \frac{|J|}{2}-
\sqrt{J^2+D^2}$. Then at temperature zero this state becomes
maximally entangled ground state. This way the DM interaction
creates entanglement in the ferromagnetic case. With increasing $D$
the gap between singlet ground state and the first doublet state
increases, this is why entanglement in the ferromagnetic case
increases.

\section{$XXZ$ Heisenberg Model} When $J_x=J_y=J \neq J_z$ the Hamiltonian (\ref{Hamiltonian1}) becomes \be
H= \frac{1}{2}[ J (\sigma^x_1 \sigma^x_2 + \sigma^y_1 \sigma^y_2 +
\Delta \, \sigma^z_1 \sigma^z_2) +B_+\, \sigma^z_1 + B_-\,
\sigma^z_2 + D(\sigma^x_1 \sigma^y_2- \sigma^y_1 \sigma^x_2 )]. \ee
where $\Delta \equiv {J_z}/{J}$.

\begin{itemize}

\item In a pure $XXZ$ ferromagnetic model when $J_z<0$ and
$-|J_z|<J<|J_z|$ or $|\Delta|>1$, we have the degenerate maximal
eigenvalues  $\lambda_1=\lambda_2$ and no entanglement occurs. This
happens since the ground state of the system is doublet with
eigenvalues $E_1=E_2= -\frac{|J_z|}{2}$. \item In particular case
$|\Delta|=1$ or $|J|= |J_z|$ we have reduction to the $XXX$ model,
where the energy level $E_3$ merges to the ground state, and the
last one becomes triplet state, as we discussed above in Sec.6.
\item For $J>0$ and $\Delta>-1$ the maximal eigenvalue is
$\lambda_3$ and the states are entangled when
 $\sinh \frac{J}{kT}>
 e^{-J_z/kT}$ with the concurrence
 {\small \be C_{12}= \frac{\sinh \frac{J}{kT}- e^{-J_z/kT}}{
\cosh \frac{J}{kT}+ e^{-J_z/kT}} \label{concpureXXZ}.\ee} \item For
$J<0$ and $\Delta<1$ the maximal eigenvalue is $\lambda_4$
 and the states are entangled for
$\sinh \frac{|J|}{kT}>
 e^{-J_z/kT}$ with the concurrence
{\small \be C_{12}= \frac{\sinh \frac{|J|}{kT}- e^{-J_z/kT}}{ \cosh
\frac{|J|}{kT}+ e^{-J_z/kT}} \label{concpureXXZ}.\ee}
\end{itemize}
\subsection{$XXZ$ Heisenberg Model with DM Coupling ($B =0, b = 0, D\neq 0$)}
With addition of the DM coupling we have the eigenvalues \be
\lambda_{1,2}= \frac{1}{2\left[1+ e^{J_z/kT} \cosh
\frac{\sqrt{J^2+D^2}}{kT}\right]},\,\,\, \lambda_{3,4}= \frac{e^{\mp
\sqrt{J^2+D^2}/kT}}{2\left[e^{-J_z/kT}+ \cosh
\frac{\sqrt{J^2+D^2}}{kT}\right]}.\ee Then for $J_z<0$ and $|J_z|>
|J|$, there exists critical value $D_c= \sqrt{J_z^2-J^2}$ so that
for $D>D_c$ and $\sinh \frac{\sqrt{J^2+D^2}}{kT}> e^{-J_z/kT}$ the
states are entangled with the concurrence {\small \be C_{12}=
\frac{\sinh \frac{\sqrt{J^2+D^2}}{kT}- e^{|J_z|/kT}}{ \cosh
\frac{\sqrt{J^2+D^2}}{kT}+e^{|J_z|/kT} }. \label{concpureXXZDM1}\ee
} This happens because for $J_z<0$, $|J_z|> |J|$ and $D=0$, the
ground state is doublet with $E_1=E_2= -\frac{|J_z|}{2}$, and by
increasing $D$ so that $D>D_c$, the higher energy level $E_3$ lowers
to the singlet ground state which is maximally entangled. Comparison
of (\ref{concpureXXZDM1}) with (\ref{concpureXXZ}) shows that with
 growing $D$ entanglement increases.

 It is worth to note that the concurrence
(\ref{concpureXXZDM1})
 for both signs of $J$ is the same.
Moreover, as easy to see in (\ref{concpureXXZDM1}) parameters $J$
and $D$ appear symmetrically. It means that the concurrence could be
increased by growing $J$ with fixed $D$ either by growing $D$ with
fixed $J$. This reflects the known result \cite{alcaraz} on
equivalence of the Heisenberg $XXZ$ model with DM coupling to pure
$XXZ$ model with modified anisotropy parameter and a certain type of
boundary conditions. In fact comparing entanglement in our formulas
for pure antiferromagnetic case (\ref{concpureXXZ}) with the one
including the DM interaction (\ref{concpureXXZDM1}), we can see that
the concurrences are connected by the replacement $J \rightarrow J
\sqrt{ 1+ \frac{D^2}{J^2}}$, which corresponds to the substitution
for the anisotropy parameter in the pure $XXZ$ model as $\Delta
\rightarrow \frac{\Delta}{\sqrt{1+ \frac{\Delta^2 D^2}{J_z ^2}}}$.

\subsection{$XXZ$ Heisenberg Model with DM Coupling and Magnetic Field}
If we take into account the DM interaction $D$ and magnetic field
$B$ simultaneously, the above results for critical value of the DM
coupling are still valid, but the level of entanglement decreases
according to {\small \be C_{12}= \frac{\sinh
\frac{\sqrt{J^2+D^2}}{kT}- e^{-J_z/kT}}{ \cosh
\frac{\sqrt{J^2+D^2}}{kT}+e^{-J_z/kT} \cosh{\frac{B}{kT}} }.
\label{concpureXXZDM}\ee }
  For
$T=0$ and $J_z>0$ we have nonanalytic behavior
 \be C_{12}=\left\{%
\begin{array}{ll}
  1 , & \hbox{$ \sqrt{D^2+J^2}>B-J_z$;} \\ \\
 \frac{1}{2}, & \hbox{$\sqrt{D^2+J^2}=B-J_z$;}\\ \\
  0   , & \hbox{$\sqrt{D^2+J^2}<B-J_z$.}
\end{array}%
\label{51}\right.     \ee which signals appearance of quantum phase
transitions.  The concurrence versus temperature for different
values of coupling $D$ is shown in Fig. 5, where $J=1$ , $J_z=0.5$
and magnetic field $B=2$. As we can see in general the entanglement
decreases with growing temperature. However we like to emphasize
that for $D<D_c$ in Fig. 5a, when $D=0.1$, the entanglement is
increasing with growing temperature. This phenomena can be explained
by the fact that for such values of the parameters at $T=0$ the
ground state is the separable state with energy $E_1=
\frac{J_z}{2}-B= -1.75$, and the concurrence is zero (see the last
case in eqn. (\ref{51})). When temperature increases the entangled
state with energy $E_3= \frac{-J_z}{2}\mp \sqrt{J^2+D^2}= -1.255$
becomes involved into the mixture and entanglement is increasing.

When $D=D_c$ the entanglement decreases smoothly from $C_{12}=0.5$
(Fig. 5b, $D_c=1.118$). By increasing $D$ ($D=1.19$), first it gives
sharp decrease from $C_{12}=1$ (Fig. 5c) and then it vanishes
slowly. When $D$ becomes bigger ($D=3$) entanglement decreases
slowly from $C_{12}=1$ (Fig. 5d).
\begin{figure}[h]
\begin{center}
\epsfig{figure=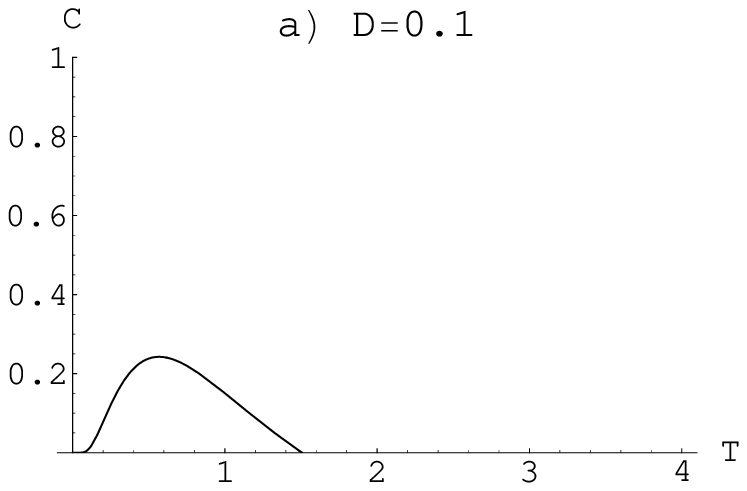,height=3cm,width=3cm}
\epsfig{figure=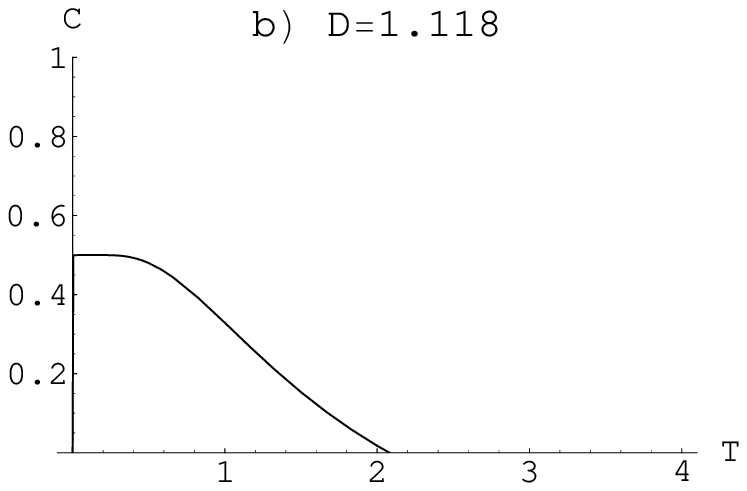,height=3cm,width=3cm}
\epsfig{figure=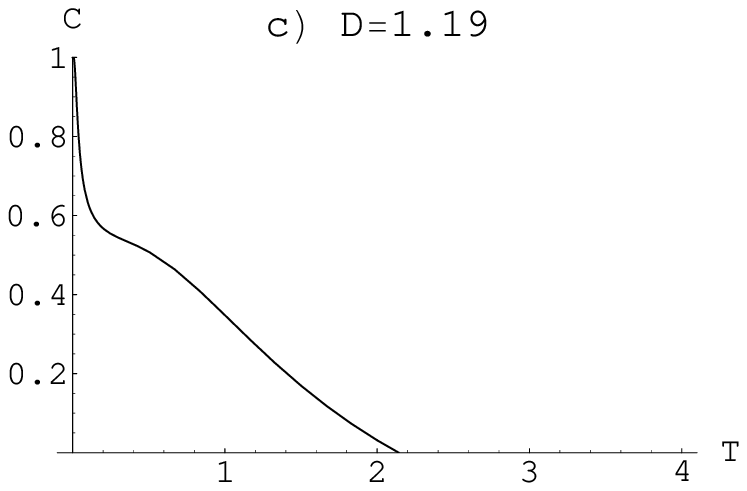,height=3cm,width=3cm}
\epsfig{figure=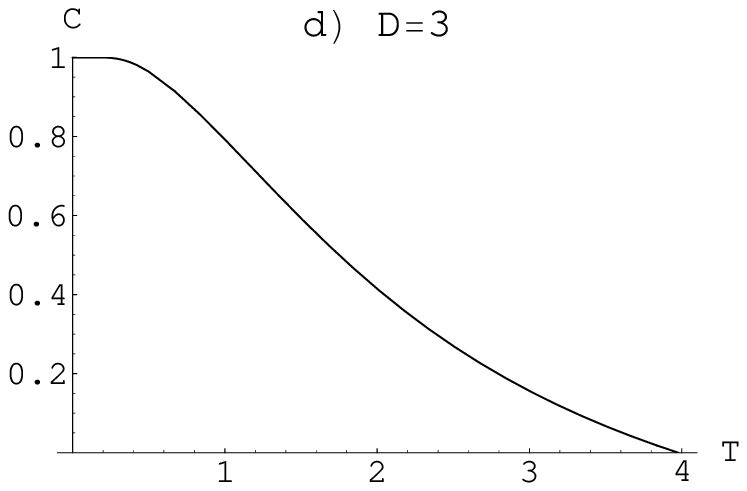,height=3cm,width=3cm}
\end{center}
\caption{Concurrence in $XXZ$ model versus temperature for $B=2$ and
a) $D=0.1$, b)$D= 1.118$, c)$D = 1.19$, d)$D = 3$}
\end{figure}

We compare the concurrence versus magnetic field for different
temperatures, when $D=0$ (Fig. 6)  and  when $D=2$ (Fig.7). In both
cases at $T=0$ the entanglement vanishes abruptly as $B$ crosses
critical value $B_c= \sqrt{B^2+J^2}+ J_z$. This special point $T=0$,
$B=B_c$ at which entanglement becomes nonanalytic function of $B$,
is the point of quantum phase transition. Comparison of figures 6
and 7 shows that the critical value of $B$ at which entanglement
disappears suddenly is growing with increasing coupling $D$: in
Fig.6, $B_c=2$ and in Fig.7, $B_c=3.3$. It shows again that
increasing DM coupling improves entanglement.

\begin{figure}[h]
\begin{center}
\epsfig{figure=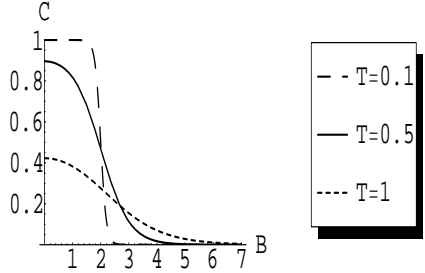,height=4cm,width=6cm}
\end{center}
\caption{Concurrence versus magnetic field $B$ for $D=0$ and $T=0.1,
0.5,1$. }
\end{figure}

\begin{figure}[h]
\begin{center}
\epsfig{figure=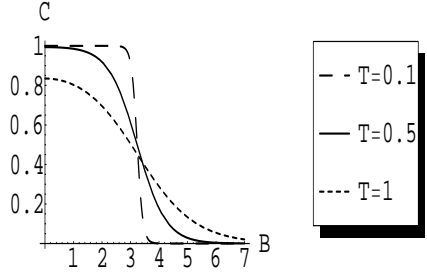,height=4cm,width=6cm}
\end{center}
\caption{Concurrence versus magnetic field $B$ for $D=2$ and $T=0.1,
0.5,1$. }
\end{figure}

%
%
%

\section{$XYZ$ Heisenberg Model}
In the present paper we are not going to analyze all possibilities
for the $XYZ$ model. Instead we restrict ourselves with a particular
range of parameters to study the influence of DM coupling in
details.

\subsection{Antiferromagnetic Case}
\subsubsection{Pure Antiferromagnetic Model}
 We start from the pure XYZ model,
where for determinacy we chose $J_z>J_y>J_x>0 $ implying $J_+>
|J_-|>0, J_- = -|J_-|<0$. Eigenstates of the Hamiltonian
(\ref{Hamiltonian1}) are $E_{1,2}= \frac{|J_z|}{2} \pm |J_-|$ and
$E_{3,4}= -\frac{|J_z|}{2} \mp |J_+|$. For zero temperature the
ground state is maximally entangled Bell state $|01\rangle -
|10\rangle$ with the energy $E_{3}= -\frac{|J_z|}{2} - |J_+|$. When
temperature increases, the state mixes with higher states decreasing
entanglement. To find concurrence we have the biggest eigenvalue
$\lambda_4= \frac{1}{Z} \exp{\frac{|J_z|+2|J_+|}{2kT}}$ and \be
C_{12}=max \{\frac{\sinh \frac{J_+}{kT}-\cosh
\frac{J_-}{kT}e^{-J_z/kT} }{\cosh \frac{J_+}{kT}+ \cosh
\frac{J_-}{kT}e^{-J_z/kT} },0\}.  \ee  Then entanglement occurs when
$\sinh \frac{J_+}{kT}>\cosh \frac{J_-}{kT}e^{-J_z/kT} $. It shows
that entanglement depends essentially on the anisotropy, and grows
with $J_+$ and decreases with $J_-$ \cite{rigolin}.

\subsubsection{$XYZ$ Model with DM Coupling} Inclusion of the DM coupling, remains the energy levels $E_1$ and $E_2$
the same as above, while $E_{3,4}= -\frac{|J_z|}{2} \mp \sqrt{J_+^2+
D^2}$. In this case the ground state continues to be entangled state
but with the energy $E_3$. With growing temperature, mixing of this
state with the higher states decreases the entanglement. If we
consider the difference between two lower states $E_4- E_3=
\sqrt{J_+^2+ D^2}$, then by increasing the coupling $D$, it can be
made arbitrary large, so that the entanglement will increase. For
$D>> |J_+|$ the state would be maximally entangled. At the
concurrence
$T=0$  \be C_{12}=\left\{%
\begin{array}{ll}
    1, & \hbox{$\sqrt{D^2+J_+^2}> J_-- J_z$;} \\
    0, & \hbox{$\sqrt{D^2+J_+^2}= J_-- J_z$;} \\
    1, & \hbox{$\sqrt{D^2+J_+^2}< J_-- J_z$,} \\
\end{array}%
\right. \ee is nonanalytic function in $D$, and it signals about the
quantum phase transition at $D=D_c$ where $\sqrt{D_c^2+J_+^2}=
J_--J_z$. When the temperature increases, entanglement occurs for
\be \sinh \frac {\sqrt{J_+^2+ D^2}}{kT}> e^{-J_z/kT}\cosh
\frac{J_-}{kT}, \ee and the concurrence  {\small \be C_{12}=
\frac{\sinh \frac {\nu}{kT}-e^{-J_z/kT} \cosh\frac{J_-}{kT} }{\cosh
\frac{\nu}{kT}+e^{-J_z/kT} \cosh \frac{J_-}{kT}}, \ee } increases
with growing anisotropy $J_+$ and the coupling $D$.
 \subsection{Ferromagnetic Case ($J_z<J_y<J_x<0$)}
\subsubsection{Pure $XYZ$ Model}
 Let
$J_z<J_y<J_x<0$ then $J_+=-|J_+|$, $J_-=|J_-|>0$ and $J_z=-|J_z|$.
For pure $XYZ$ model, eigenstates of the Hamiltonian are $E_{1,2}=
-\frac{|J_z|}{2} \mp |J_-|$ and $E_{3,4}= \frac{|J_z|}{2} \pm
|J_+|$. For zero temperature the ground state is maximally entangled
Bell state $|00\rangle - |11\rangle$ with the energy $E_{1}=
-\frac{|J_z|}{2} - |J_-|$. With increasing temperature this state
mixes with other states and entanglement decreases so that the
concurrence \be C_{12}=\frac{\sinh \frac{|J_-|}{kT}-\cosh
\frac{|J_+|}{kT}e^{-|J_z|/kT} }{\cosh \frac{|J_-|}{kT}+ \cosh
\frac{|J_+|}{kT}e^{-|J_z|/kT} }.\ee When temperature reaches the
critical value $T=T_c$, given by a solution of the following
transcendental equation \be \sinh \frac{|J_-|}{kT_c}=\cosh
\frac{|J_+|}{kT_c}e^{-|J_z|/kT},\ee the concurrence vanishes and
state becomes unentangled.

\subsubsection{$XYZ$ Model with DM Coupling}
 With inclusion of the DM coupling, the first couple of energy levels
is the same  $E_{1,2}= \frac{-|J_z|}{2} \mp |J_-|$ while the second
couple becomes $E_{3,4}= \frac{|J_z|}{2} \mp \sqrt{J_+^2+ D^2}$. For
$D<D_c$ where $D_c$ satisfies the equation $\sqrt{D_c^2+J_+^2}=
|J_z|+|J_-|$, the ground state of the system is the maximally
entangled Bell state $|00\rangle - |11\rangle$. If we increase $D$,
the difference between energy levels $E_1$ and $E_3$ decreases, so
that at $D=D_c$ the ground state becomes degenerate and entanglement
vanishes. When $D> D_c$ the ground state $E_3$ becomes entangled
again.

Due to the mixture of states by increasing temperature the
entanglement decreases, so that, in the under critical region
$D<D_c$ the concurrence is \be C_{12}= max \{\frac{\sinh
\frac{|J_-|}{kT}-\cosh \frac{\sqrt{J_+^2+D^2}}{kT}e^{-|J_z|/kT}
}{\cosh \frac{|J_-|}{kT}+ \cosh
\frac{\sqrt{J_+^2+D^2}}{kT}e^{-|J_z|/kT} }, 0 \}, \ee while in the
over critical region $D>D_c$ it is  \be C_{12}= max \{ \frac{\sinh
\frac {\sqrt{J_+^2+D^2}}{kT}-e^{|J_z|/kT} \cosh\frac{|J_-|}{kT}
}{\cosh \frac{\sqrt{J_+^2+D^2}}{kT}+e^{|J_z|/kT} \cosh
\frac{|J_-|}{kT}}, 0 \} .\ee

For $D=D_c$, due to $\lambda_1= \lambda_3$, the entanglement
vanishes for any temperature. The entanglement dependence on $T$ and
$D$ is shown in Figs.8 and 9. For $T=0$ the figures show
nonanalyticity at $D=D_c$ which signals a quantum phase transition.
The entanglement behavior in the under and the over critical regions
is qualitatively different. For the under critical case with fixed
temperature the entanglement decreases with growing $D$, and the
level of entanglement quickly decreases with temperature. From
another side, for fixed temperature in the over critical region  the
entanglement increases, and the level of entanglement decreases with
temperature quite slowly. In addition if at $T=0$ we have only one
critical point $D=D_c$ in which entanglement is zero, for $T>0$
entanglement vanishes at some interval which includes $D_c$ and this
interval extends with growing temperature. This is a result of
ground state mixture with higher states. However by increasing $D$
we can always lower the level of our ground state to decrease this
mixture and increase entanglement.

\begin{figure}[h]
\begin{center}
\epsfig{figure=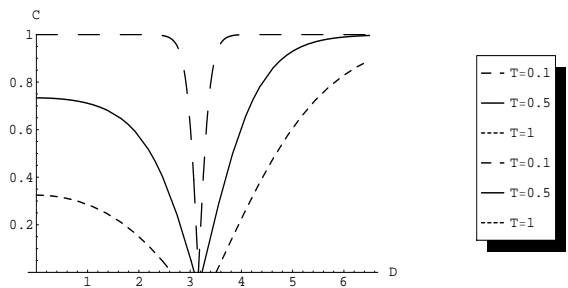,height=4cm,width=8cm}
\end{center}
\caption{Concurrence in ferromagnetic XYZ model versus  coupling $D$
at temperature $T=0.1,\, 0.5,\, 1$ }
\end{figure}

\begin{figure}[h]
\begin{center}
\epsfig{figure=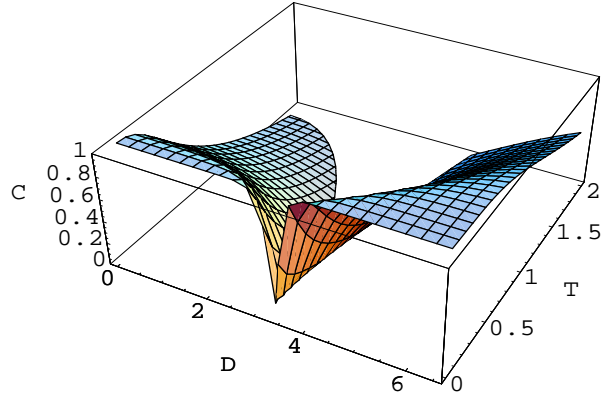,height=6cm,width=8cm}
\end{center}
\caption{3D plot of concurrence in ferromagnetic XYZ model versus
 coupling $D$ and temperature $T$ }
\end{figure}

\section{Conclusion}
In conlusion we like to stress several implications for future
studies. As was shown by Alcaraz et al. \cite{alcaraz} the $XXZ$
quantum Heisenberg chain with the DM interaction is equivalent to
the pure $XXZ$ Hamiltonian with modified boundary conditions and the
anisotropy parameter, so that with these boundary conditions the
model is still solvable by the Bethe Ansatz. Taking into account our
results it shows possibility to control entanglement in $XXZ$ model
by varying boundary conditions.

 Recently it was found that the DM interaction can excite the
entanglement and teleportation fidelity by using two independent
Heisenberg $XXX$ chains \cite{zhang}.  Moreover, studying the effect
of a phase shift on amount transferable two-spin entanglement
\cite{maruyama}, it was shown that maximum attainable entanglement
is enhanced by the DM interaction.  Very recently geometric
computations for a spin chain model with the DM interaction has been
discussed in  \cite{Jones}. Finally it was found that the DM
interaction is present in number of quasi-one dimensional magnets
and is dominating  for the compound $RbCoCl_3. 2 H_2O$.These
indicate that DM interaction could be significant in designing the
spin-based realistic quantum computers \cite{kavokin}. The above
mentioned results suggest to study the most general $XYZ$ Heisenberg
model with DM interaction as a quantum channel for quantum
teleportation. These questions are now under investigation.

{\bf Acknowledgments} \noindent  One of the authors (Z.N.G.) would
like to thank Dr. Koji Maruyama,
 for his helpful remarks.  This work was  supported partially by
Izmir Institute of Technology, Turkey.

\end{document}